\documentclass{moriond}

\def\Journal#1#2#3#4{{#1} {\bf #2}, #3 (#4)}

\def\PLB{{\em Phys. Lett.}  B}

\def\PRD{{\em Phys. Rev.} D}

\def\be{\begin{equation}}
\def\ee{\end{equation}}
\def\bea{\begin{eqnarray}}
\def\eea{\end{eqnarray}}

\newcommand{\Photo}{}

\begin{document}
\vspace*{4cm}
\title{GENUINE EXTRA YUKAWAS FROM EXTRA HIGGS, \ IMPLICATIONS}

\author{ GEORGE WEI-SHU HOU }

\address{Department of Physics, National Taiwan University, %\\
Taipei 10617, Taiwan}

\maketitle\abstracts{
With a second Higgs doublet, extra Yukawa couplings 
$\rho_{ij}$ generally exist.
Baryon Asymmetry of the Universe (BAU) can be accounted for
by $\rho_{tt} \sim {\cal O}(1)$, 
with first order electroweak phase transition (EWPT) 
arising from ${\cal O}(1)$ Higgs quartic couplings.
The latter can explain why the observed
$h(125)$ boson so resembles the Standard Model (SM) Higgs: 
with coupling $\eta_6 \sim{\cal O}(1)$ for two-doublet mixing, 
the $H$--$h$ mixing angle $\cos\gamma \cong -\eta_6 v^2/(m_H^2 - m_h^2)$
is suppressed by the $CP$-even boson mass splitting $m_H^2 - m_h^2 > {\rm few}\ v^2$.
The approximate alignment, together with the fermion mass-mixing pattern,
controls FCNC Higgs effects at low energy.
The picture can be probed by
 $pp \to tt\bar c$, $tt\bar t$, i.e. same-sign top and triple-top 
processes at the LHC.
}

\section{Introduction: Whither Extra Yukawas?}

Though accounting for all observed $CP$ violation (CPV), 
the unique phase in CKM matrix falls far short of BAU.
%or absence of antimatter in our Universe.
%
Considering the origin of this phase,
{\it could there be extra Yukawa couplings?}
In general, a second Higgs doublet (2HDM) --- quite plausible ---
should imply extra Yukawas, but these were killed~\cite{gw} by the 
{\it Natural} Flavor Conservation (NFC) condition:
as $u$- and $d$-type quark masses each arise from a single doublet,
the Yukawa couplings are basically the same as in SM.
It was later noted that the fermion  mass-mixing pattern 
could soften the need for NFC, and  the best probe~\cite{tch} may be
$t\to ch$ or $h \to t\bar c$, as the top quark is the heaviest fermion.

With the case for 2HDM elevated by 
the discovery of 125 GeV boson in 2012,
we emphasized~\cite{chkk} the need to probe the $2\times 2$ 
{\it extra Yukawa couplings} $\rho_{ij}$ ($i,\,j = c,\,t$).
It  also became understood that the flavor changing neutral Higgs (FCNH) couplings 
of the form
\begin{equation}
\rho_{tc} \cos(\beta-\alpha)\, \bar t_Lc_Rh,
\end{equation}
are modulated by $H$--$h$ mixing, where $H$ is the second
$CP$-even Higgs boson.
The two doublets $\Phi_1$ and $\Phi_2$
give rise to Yukawa matrices ${\boldmath Y}_1$ and ${\boldmath Y}_2$.
The combination ${\boldmath Y}^{\rm SM} = {\boldmath Y}_1v_1 + {\boldmath Y}_2v_2$ 
is diagonalized as usual, but the orthogonal combination gives rise to 
Yukawa matrix {\boldmath $\rho$}
that cannot be simultaneous  diagonalized.
In the limit that $\cos(\beta-\alpha)$ is small, called {\it alignment} limit,~\cite{gunhab} 
couplings of $h$ are diagonal, just as the SM Higgs, while $H$ couples with 
the Yukawa matrix {\boldmath $\rho$}.
As Yukawa couplings, $\rho_{ij}$ should be complex, 
\begin{equation}
\rho_{ij} \equiv |\rho_{ij}|e^{i\phi_{ij}}.
\end{equation}
In place of NFC, we see that
alignment ($\cos(\beta-\alpha) \to 0$) removes FCNH couplings for $h$,
while the mass-mixing pattern, shared by $Y_{1,2}$, 
further suppresses $\rho_{ij}$ involving light(er) quarks.

\section{\boldmath Bonus 1: EWBG from \underline{Extra} Top Yukawa $\rho_{tt}$}

Given that $Y^{\rm SM}$ for $u$-type quarks is dominated by
Yukawa coupling $\lambda_t \cong 1$, together with
the observed quark mass-mixing pattern, it is rather plausible that 
the orthogonal combination to $Y^{\rm SM}$ should also have 
a dominant ${\cal O}(1)$ eigenvalue, with phase arbitrary.
This motivates us~\cite{fhs} to consider its possible role in baryogenesis.
It is known that~\cite{kos} thermal loops involving extra Higgs bosons
with  ${\cal O}(1)$ Higgs quartic couplings can give rise to
1st order EWPT. It is of interest to explore whether Im$\,\rho_{tt}$ could
then lead to electroweak baryogenesis (EWBG).

The main issue is to generate sufficient $Y_B \equiv n_B/s$ 
(ratio of baryon and entropy densities) at the observed 
level of $Y_B^{\rm obs} \sim 0.86 \times 10^{-10}$ or higher. 
Putting aside the complicated transport problem,~\cite{fhs} 
which requires an actual 1st order EWPT, 
this boils down to producing enough left-handed top density at the expanding 
bubble wall of broken phase that accumulates inside the bubble, i.e. our Universe.
This depends on CPV top interactions at the bubble wall,
which boils down further to the CPV source term that arises from 
the extra top Yukawas,~\cite{fhs}
\begin{equation}
{\rm Im}\big[(Y_1)_{ij}(Y_2)_{ij}^*\big]
= {\rm Im}\big[(V_L^uY_{\rm diag} V_R^{u\dagger})_{ij}(V_L^u\rho V_R^{u\dagger})_{ij}^* \big],
\label{eq:YCPV}
\end{equation}
where $V_L^u$, $V_R^u$ forms the biunitary transform that
diagonalizes  $Y^{\rm SM}$ to $Y_{\rm diag}$ for $u$-type quarks.

\begin{figure}
\hskip1.3cm 
\begin{minipage}{0.35\linewidth}
\centerline{\includegraphics[width=1.1\linewidth]{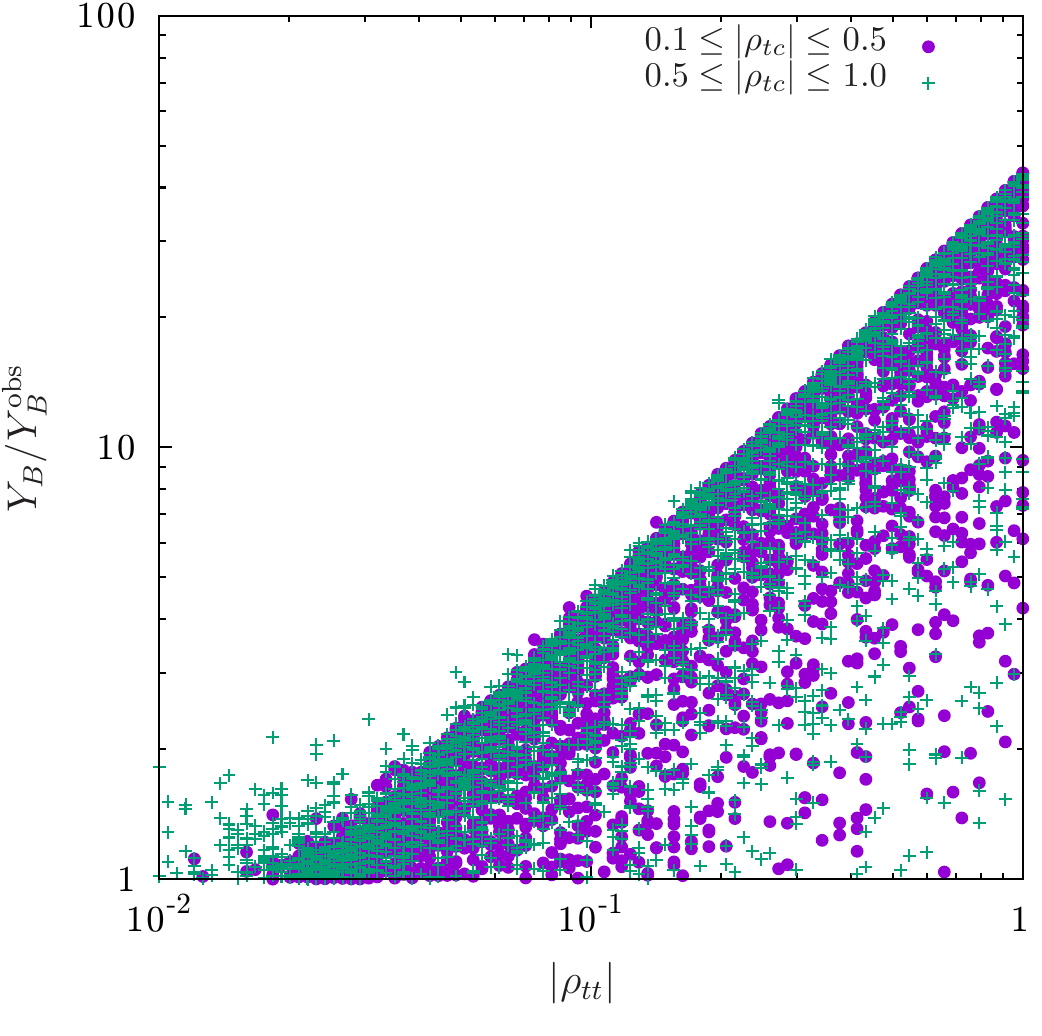}}
\end{minipage}
\hskip1.75cm %\hfill
\begin{minipage}{0.33\linewidth}
\centerline{\includegraphics[width=1.25\linewidth]{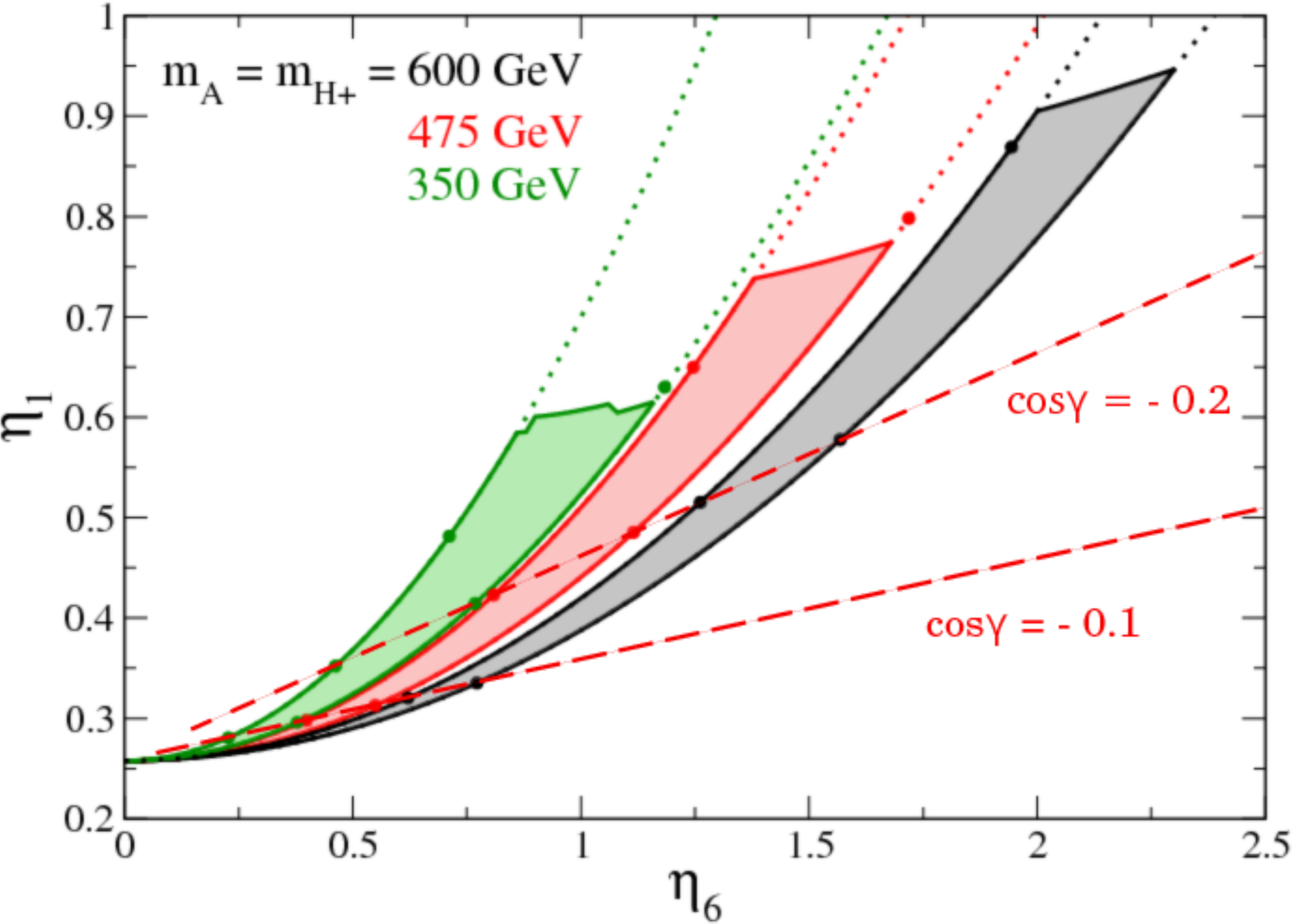}}
\end{minipage}
\caption[]{
 [left] $Y_B/Y_B^{\rm obs}$ vs $|\rho_{tt}|$, and
 [right] $\Delta T$-allowed $\eta_1$ vs $\eta_6$ for $m_A = m_{H^+}$,
varying $\eta_4 = \eta_5 \in (0.5,\,2)$.
}
\label{fig:radish}
\end{figure}

Flavor constraints from $B_d$ and $B_s$ mixing and chiral enhancement in 
$b\to s\gamma$ demand~\cite{chkk,ahkkm} $\rho_{ct}$ to be rather small, 
while $\rho_{cc} \sim {\cal O}(\lambda_c) \ll 1$ without fine tuning, 
hence the two main parameters are $\rho_{tt}$ and $\rho_{tc}$.
Scanning over $|\rho_{tc}|$, $\phi_{tc}$ and $\phi_{tt}$, 
we find robust and large parameter space for EWBG.
Fig.~1[left] plots $Y_B/Y_B^{\rm obs}$ vs $ |\rho_{tt}| \in (0.01,\, 1)$,
with higher $0.5 \leq |\rho_{tc}| \leq 1.0$ (lower $0.1 \leq |\rho_{tc}| \leq 0.5$) 
plotted as green $+$ (purple {\boldmath  $\cdot$}).
Little difference is seen between the two plots, hence $\rho_{tt}$ is the driver.
However, for $|\rho_{tt}| < 0.05$ or so, 
the green  $+$'s that populate $Y_B/Y_B^{\rm obs} > 1$ 
suggest $\rho_{tc} > 0.5$, with phase $\phi_{tc}$ near maximal,
could be a backup to $\rho_{tt}$ for EWBG.
In making this plot, the simplifying assumption of
$m_H = m_A = m_{H^+} = 500$ GeV is taken.
Much higher values would either run into issues of perturbativity,
or damping by decoupling.

Fig.~1[left] scanned through realistic Yukawa matrices,
but a simplified texture can help elucidate the driving effect.
Suppose
 $(Y_1)_{tc} \neq 0,\, (Y_2)_{tc} \neq 0$ and
 $(Y_1)_{tt}= (Y_2)_{tt} \neq 0$, 
while all other extra Yukawas vanish, i.e. altogether 3 complex parameters.
If one assumes $\sqrt2 Y^{\rm SM}$ is the linear sum of $Y_1$ and $Y_2$,
one can solve for $V^u_R$, while there is no need for $V^u_L$.
One can then arrive at the combination of $Y_1$ and $Y_2$ that is
orthogonal to $Y^{\rm SM}$. In this way, one finds~\cite{fhs}
\begin{equation}
 {\rm Im}\big[ (Y_1)_{tc}(Y_2)_{tc}^* \big]
 = -\lambda_t\, {\rm Im}\,\rho_{tt}, \quad \rho_{ct}=0,
\label{eq:simpCPV}
\end{equation}
with $\rho_{tc}$ remaining basically a free parameter.
We see from Eq.~(\ref{eq:simpCPV}) that {\it both} doublets participate
in the CPV source for EWBG in 2HDM, which is 
reminiscent to the Jarlskog invariant for SM.
We can also see how 2HDM with extra Yukawas overcomes 
the suppression factors in the Jarlskog invariant, given that
$\lambda_t$, $|{\rm Im}\,\rho_{tt}|$ are both ${\cal O}(1)$.

\section{\boldmath Bonus 2: :  Alignment from ${\cal O}(1)$ Higgs Quartics}

{\it It is remarkable that the extra Yukawa coupling 
$\rho_{tt}$ could account for BAU!} 

Note that such mechanism does not exist in
2HDM-I or 2HDM-II, the 2HDMs that satisfy NFC, since NFC
means there are essentially no new Yukawa couplings, 
despite having a second Higgs doublet.
We now show~\cite{hk2} that the prerequisite for 1st order EWPT, that
extra Higgs quartic couplings are ${\cal O}(1)$, could be behind
the observed approximate alignment.

The general $CP$-conserving Higgs potential~\cite{haber} of 2HDM is, 
\bea
 V(\Phi,\, \Phi')
&= \mu_{11}^2 |\Phi|^2 +\mu_{22}^2 |\Phi'|^2
        - \left(\mu_{12}^2 \Phi^\dagger \Phi' + {\rm h.c.}\right)   %\nonumber \\
   \ +\frac{1}{2}\eta_1^{}|\Phi|^4 + \frac{1}{2}\eta_2^{}|\Phi'|^4 + \eta_3^{}|\Phi|^2|\Phi'|^2  \nonumber \\
   & + \eta_4^{}|\Phi^\dagger \Phi'|^2
   + \left\{\frac{1}{2}\eta_5^{}(\Phi^\dagger \Phi')^2
   + \left[ \eta_6^{}|\Phi|^2 + \eta_7^{}|\Phi'|^2\right] \Phi^\dagger \Phi' + {\rm h.c.}\right\},
 %\notag  %
 \label{eq:pote2}
\eea
where we take Higgs basis, i.e.  $\mu_{11}^2 < 0$ but $\mu_{22}^2 >0$.
With the two minimization conditions
\begin{equation}
  \mu_{11}^2 = -\frac{1}{2}\eta_1^{}v^2, \quad
  \mu_{12}^2 =  \frac{1}{2}\eta_6^{}v^2,
 \label{eq:minim}
\end{equation}
$\mu_{11}^2 < 0$ is exchanged for $v$, the usual
``soft breaking parameter'' $\mu_{12}^2$ is removed, and 
the quartic coupling $\eta_6$ is solely responsible for $\Phi$--$\Phi'$ mixing.
The $CP$-even Higgs mass matrix 
\begin{equation}
  M_\textrm{even}^2 =
  \left[\begin{array}{cc}
    \eta_1^{}v^2 & \eta_6^{} v^2 \\
    \eta_6^{} v^2 & \mu_{22}^2 + \frac{1}{2} (\eta_3^{} + \eta_4^{} + \eta_5^{})v^2\\
    \end{array}\right], \quad\quad
%  \,\,\,\,\,
  R_\gamma  =  \left[\begin{array}{cc}
    c_\gamma & - s_\gamma \\
    s_\gamma & c_\gamma \\
  \end{array}\right],
\end{equation}
is diagonalized by $R_\gamma$, 
i.e. $R^T_\gamma M_\textrm{even}^2 R_\gamma$ is diagonal 
with elements $m_H^2$, $m_h^2$.
In Eq.~(7), our $c_\gamma \equiv \cos\gamma$ corresponds to 
$\cos(\beta - \alpha)$ in the 2HDM-II notation, and is the relative angle (mod. $\pi/2$) 
between the Higgs basis and the neutral Higgs mass basis.

Rather than give the formula for $m_H^2$, we note the mixing angle 
$c_\gamma$ satisfies two relations,
\begin{equation}
   c_\gamma^2  = \frac{\eta_1^{}{v^2} - m_h^2}{m_H^2 - m_h^2}, \quad \ \  %\label{eq:npara1}\\
  \sin2\gamma^{}  = \frac{2\eta_6^{}v^2}{m_H^2 - m_h^2}.
 \label{eq:npara}
\end{equation}
In alignment \emph{limit} of $c_\gamma \to 0$, $s_\gamma \to -1$,
one has $\eta_1 \to m_h^2/v^2 \simeq 0.26$ in numerator of first term, 
where $m_h \simeq 125$ GeV is used.
For $c_\gamma$ small but nonvanishing, 
$m_H^2 - m_h^2 >$ several $v^2$ can weigh down  
$|\eta_1 v^2 - m_h^2| < v^2$.
Since $s_\gamma \to -1$ holds better than $c_\gamma \to 0$, 
the second relation gives
\begin{equation}
  c_\gamma^{}  \simeq \frac{-\eta_6^{}v^2}{m_H^2 - m_h^2}.
  \quad\quad \ \ {\rm (near\; alignment)}
 \label{eq:Haber}
\end{equation}
Although the result exists~\cite{bechtle} in the literature, 
$|\eta_6| \ll 1$ is generally assumed, 
as it arises through loop effects in MSSM.
But we see that $c_\gamma$ can be small for
\begin{equation}
  |\eta_6| \sim {\cal O}(1)\; ({\rm or\ smaller}), \quad m_H^2 - m_h^2 > {\it several}\; v^2. 
 \label{eq:alignment}
\end{equation}
Note that a low $m_h^2/v^2 \simeq 0.26$ is \emph{not} required,
i.e. $c_\gamma$ can be small even if $m_h \sim 300$ GeV.

What drives alignment in 2HDM?
For $\eta_{1,3,4,5}$, $\mu_{22}^2/v^2 \sim {\cal O}(1)$,
$[M_{\rm even}^2]_{22}$ has four ${\cal O}(v^2)$ terms
while $[M_{\rm even}^2]_{11}$ has only one,
hence $m_H^2 - m_h^2 > {\rm several}\; v^2$ is likely.
However, $\mu_{22}^2/v^2 > 1$ would damp the 
%strength, and possibility, of 
1st order EWPT, hence sub-TeV exotic Higgs masses are preferred.
%, and is a boon for LHC.
Second, $\eta_6 \sim {\cal O}(1)$
increases $m_H^2 - m_h^2$ by level repulsion,
pushing $m_h^2/v^2$ down from $\eta_1 \sim {\cal O}(1)$.
Finally, tuning $\eta_6 < 1/4  \sim m_h^2/v^2$ would give 
\emph{extreme} alignment ($c_\gamma \to 0$) hence $\eta_1 \to 0.26$.
These observations are illustrated in Fig.~1[right] for 
allowed $\eta_1$ vs $\eta_6$ range,
where custodial SU(2) is assumed to evade $\Delta T$ constraint, 
i.e. $m_A^2 = m_{H^+}^2 = \mu_{22}^2 + \eta_3 v^2/2$.
We vary $\eta_4 = \eta_5 \in (0.5,\, 2)$, 
so $m_H$ could be up to 100 GeV higher.
High values of $\eta_1$ are cut off by $\Delta T$
(via scalar--vector loop),
and the two dashed lines mark $-c_\gamma = 0.1$ and $0.2$,
which are quite close to alignment; even $-c_\gamma = 0.3$, 
close to the bound from $\Delta T$, is still allowed by 
observed approximate alignment at LHC. 

{\it ${\cal O}(1)$ Higgs quartics could be behind 
approximate alignment}, or small $c_\gamma$, regardless of
whether a $Z_2$ symmetry is used to enforce NFC or not,
as our discussion is general.
But we have advocated that $\rho_{tt} \sim {\cal O}(1)$ could explain BAU.
It is then intriguing to comment that sizable $\rho_{tt}$
could possibly~\cite{hk} help ``protect'' alignment:
with ${\cal O}(1)$ Higgs quartics, bosonic loops
would reduce $\Gamma_{h \to ZZ^*}$, but
the top loop can bring $\Gamma_{h \to ZZ^*}$ back to SM value 
for $\rho_{tt}\, c_\gamma > 0$, consistent with what is observed.
This was our original motivation to understand the mechanism of alignment.

\section{\boldmath Same-sign Top and Triple-top Signatures: $pp \to tH/A \to tt\bar c$, $tt\bar t$}

The process $cg \to tA$ was suggested long time ago as a direct probe~\cite{hlmy}
of the $ctA$ FCNH coupling, restricting to $m_A < 2m_t$ such that
$A \to t\bar c$ (and $\bar t c$) is at 100\%.
We recently studied~\cite{kmh} the $cg \to tH/A$ associated production 
through the $\rho_{tc}$ coupling, followed by subsequent decay 
$H/A \to t\bar c$, $\bar tc$ and $t\bar t$ final states involving 
$\rho_{tc}$ and $\rho_{tt}$ couplings, advocating 
the signatures of same-sign top, $tt\bar c$, and triple-top, $tt\bar t$.
The same-sign top signature involves same-sign dileptons, together with
two $b$-jets, missing energy, and additional jets. We find that,
for $\rho_{tc} \sim 1$, the second case for EWBG can
be probed with 300 fb$^{-1}$, but signature does not improve
for higher luminosity, unless background can be further controlled.
Given that $\rho_{tt}$ is the favored driver for EWBG, 
triple-top search at HL-LHC may be more interesting, and
possesses more exquisite signatures:
three leptons, three $b$-jets, missing energy.
The backdrop of SM cross section at only fb level makes the case strong,
where full HL-LHC data can cover up to 700 GeV mass range 
for $\rho_{tt} \sim 1$, but $\rho_{tc}$ needs to be 
not much smaller than 0.5 for signal cross section.

\section{Conclusion: $H^0$, $A^0$, $H^\pm$ in Our Time }

With ${\cal O}(1)$ Higgs quartics for 1st order EWPT, 
the extra Yukawa $\rho_{tt}$ (or $\rho_{tc}) \sim {\cal O}(1)$
 in general 2HDM is remarkably efficient for EWBG.
The ${\cal O}(1)$ Higgs quartics support approximate alignment, 
and together with quark mass-mixing hierarchy 
control low energy FCNH effect, without need for NFC.
Having $H$, $A$ and $H^\pm$ sub-TeV in mass would be a boon to LHC search,
the discovery of which in $tt\bar c$, $tt\bar t$ final states would
touch upon Matter Asymmetry of the Universe.

\section*{Acknowledgments}

We  thank our collaborators on three consecutive papers presented here
for an enjoyable 2017, and we are grateful to illuminating discussions 
with Howie Haber at Moriond QCD 2018.

\end{document}